\documentclass[12pt,preprint]{aastex}

\shorttitle{IRFM temperatures of planetary hosts}

\shortauthors{I. Ram\'{\i}rez \and J. Mel\'endez}

\newcommand{\teff}{T_\mathrm{eff}}
\newcommand{\feh}{\mathrm{[Fe/H]}}
\newcommand{\fbol}{f_\mathrm{bol}}

\begin{document}

\title{Cooler and bigger than thought?\\
Planetary host stellar parameters from the InfraRed Flux Method}

\author{Iv\'an Ram\'{\i}rez\altaffilmark{1}}
\affil{Department of Astronomy, The University of Texas at Austin,
RLM 15.202A, TX 78712-1083}\email{ivan@astro.as.utexas.edu} \and
\author{Jorge Mel\'endez\altaffilmark{1}} \affil{Department of
Astronomy, California Institute of Technology, MC 105--24,
Pasadena, CA 91125} \email{jorge@astro.caltech.edu}

\altaffiltext{1}{Affiliated to the Seminario Permanente de
Astronom\'{\i}a y Ciencias Espaciales of the Universidad Nacional
Mayor de San Marcos, Per\'u}

\begin{abstract}
Effective temperatures and radii for 92 planet-hosting stars as
determined from the InfraRed Flux Method (IRFM) are presented and
compared with those given by other authors using different
approaches. The IRFM temperatures we have derived are
systematically lower than those determined from the spectroscopic
condition of excitation equilibrium, the mean difference being as
large as 110~K. They are, however, consistent with previous IRFM
studies and with the colors derived from Kurucz and MARCS model
atmospheres. Comparison with direct measurements of stellar
diameters for 7 dwarf stars, which approximately cover the range
of temperatures of the planet-hosting stars, suggest that the IRFM
radii and temperatures are reliable in an absolute scale. A better
understanding of the fundamental properties of the stars with
planets will be achieved once this discrepancy between the IRFM
and the spectroscopic temperature scales is resolved.

\end{abstract}


\section{Introduction}

Accurate determination of the fundamental stellar parameters for
the planetary host stars is required to improve our knowledge of
the physical properties of the extrasolar planets. The
temperatures and luminosities of these stars are used to obtain
their masses and radii, which are the stellar parameters required
to determine the masses and radii of the planets from radial
velocity and transit observations, respectively.

The temperatures of the planetary host stars have been
spectroscopically determined from the excitation equilibrium of Fe
I lines (e.g. Heiter \& Luck 2003; Gonzalez et al. 2001; Laws et al. 2003; Santos,
Israelian, \& Mayor 2004). The most complete study of this kind is
that of Santos et~al. (2004, SIM04), who determined temperatures
of 98 of them. Using a different approach, Ribas et~al. (2003,
R03) determined the temperatures of these stars from infrared
photometry. Both results are similar and, in fact, consistent with
most of the previous studies, with the exception of the Fe~I
temperatures of Sadakane et~al. (2002) and Takeda et~al. (2002),
which are about 100~K lower.

To date, it has not been verified whether the spectroscopic
temperatures are consistent with those determined from stellar
angular diameters. This is basically due to the lack of dwarf
stars with measured angular diameters. However, recent
interferometric measurements (Kervella et~al. 2003, 2004; Lane,
Boden, \& Kulkarni 2001; Pijpers et~al. 2003) and transit
observations (Brown et~al. 2001) allow one to test this now. Here
we show that our IRFM temperatures are consistent with direct
measurements of angular diameters. Then we apply the IRFM to
obtain the temperatures of 92 of the known planet-hosting stars
and compare them with those obtained by other methods.

\section{Determination of temperatures and radii}

Our work is based on the InfraRed Flux Method, which compares the
quotient between the observed bolometric flux and the flux at a
given wavelength in the infrared with the quotient between
$\sigma\teff^4$ and the emergent flux in the infrared predicted by
models to get the effective temperature and angular diameter of a
star (Blackwell, Petford, \& Shallis 1980).

The stellar atmospheric parameters required for our implementation
of the IRFM have been obtained from an updated version of the
Cayrel de Strobel, Soubiran, \& Ralite (2001) catalogue and from
SIM04. The good agreement between common $\feh$ values in these
two lists allowed us to use them together. The bolometric fluxes
were calculated from the calibration of Alonso, Arribas, \&
Mart\'{\i}nez-Roger (1995, AAM95) and the photometry was taken
mainly from 2MASS. Spectral energy distributions from Kurucz
models were adopted.

The 2MASS photometry needed to be transformed into TCS colors
given the differences found between the various infrared systems
(Carpenter 2001). From a sample of main sequence stars with TCS
photometry in Alonso, Arribas, \& Mart\'{\i}nez-Roger (1994b) and
reliable 2MASS photometry we found the following relations:
\begin{mathletters}
\begin{eqnarray}
J_\mathrm{TCS}=J_\mathrm{2MASS}+0.001-0.049(J-K)_\mathrm{2MASS}\\
H_\mathrm{TCS}=H_\mathrm{2MASS}-0.018+0.003(J-K)_\mathrm{2MASS}\\
K_\mathrm{TCS} = K_\mathrm{2MASS}-0.014+0.034(J-K)_\mathrm{2MASS}
\end{eqnarray}
\end{mathletters}
with standard deviations of $\sigma=0.038$ ($N=104$),
$\sigma=0.039$ ($N=103$) and $\sigma=0.035$ ($N=107$),
respectively ($N$ is the number of stars in every fit). The
brightest planet-hosting stars are not included in our study given
the large errors associated to their 2MASS photometry. However, we
were able to calculate the temperatures of 5 bright stars
(HD~009826, HD~022049, HD~027442, HD~117176, and HD~137759) from
the Johnson photometry available in the General Catalogue of
Photometric Data (Mermilliod, Mermilliod, \& Hauck 1997, GCPD).

Application of the IRFM results in three temperatures, one for
each band: $T_J$, $T_H$ and $T_K$. In general, they were weighted
averaged with the inverse of their errors to get the final
$\teff$, that is, $\teff=(\sum_i T_i/\Delta T_i)/(\sum_i 1/\Delta
T_i)$; with $i=J,H,K$. The internal error in $\teff$ was then
calculated from $\Delta\teff=3/(\sum_i 1/\Delta T_i)$. 
Further details on our IRFM implementation will be given in a
forthcoming paper (Ram\'{\i}rez \& Mel\'endez, in preparation).

\begin{figure}
\plotone{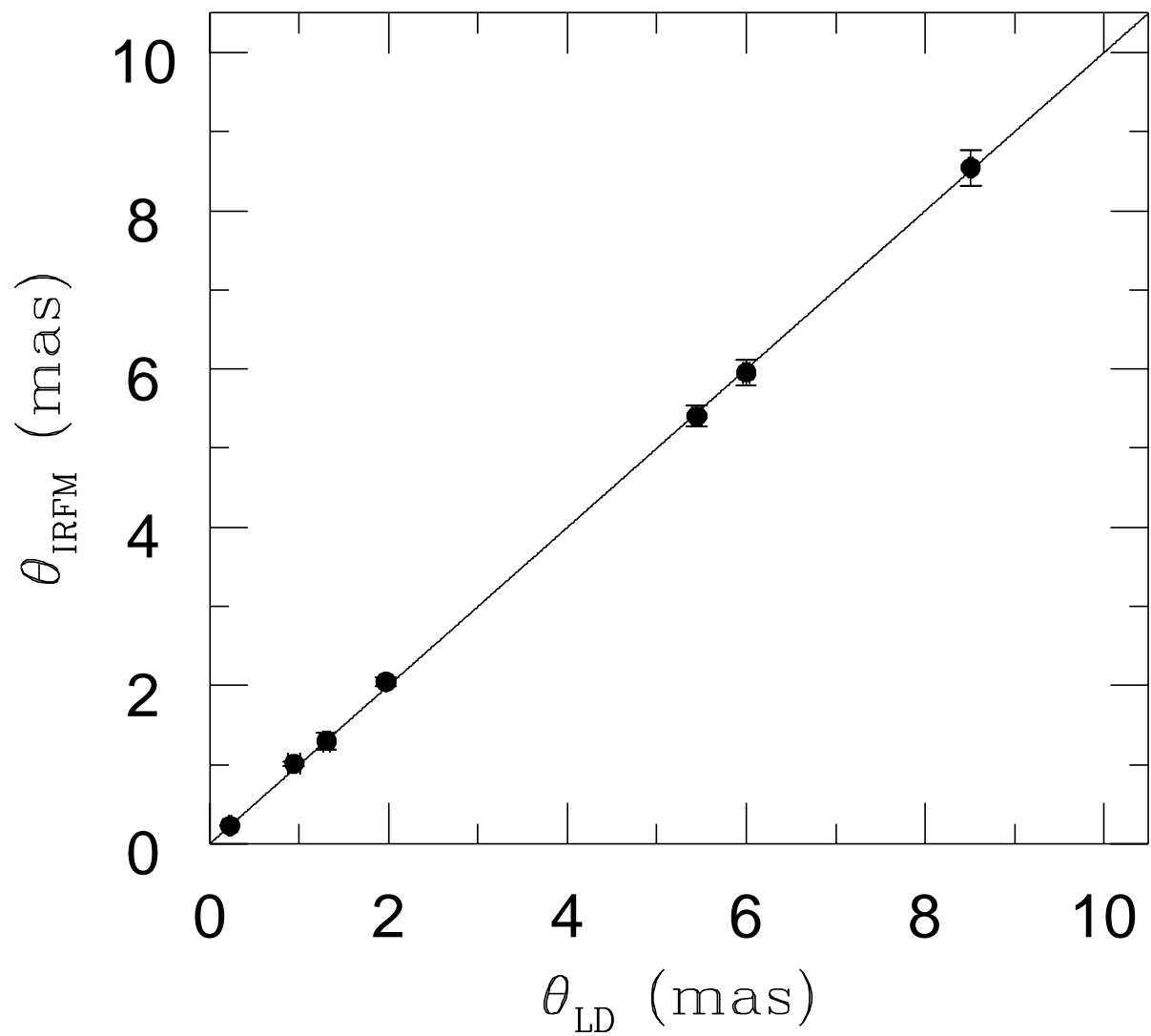} \caption{Comparison between measured angular
diameters ($\theta_\mathrm{LD}$) and diameters from the IRFM
($\theta_\mathrm{IRFM}$).} \label{fig:diams}
\end{figure}

One of the critical ingredients of the IRFM is the absolute
infrared flux calibration of the standard star. We have adopted
that given by Alonso, Arribas, \& Mart\'{\i}nez-Roger (1994a),
which was derived by demanding that the IRFM temperatures (angular
diameters) be well scaled to the direct (measured) ones, at least
for giants. This seems to be also true for dwarfs, as it is shown
in Fig.~\ref{fig:diams} and Table~\ref{table:diams}, where the
angular diameters of 7 dwarf stars derived in our work are
compared to those measured by Kervella et~al. (2003, 2004), Lane
et~al. (2001), and Pijpers et~al. (2003) from interferometric
observations and by Brown et~al. (2001) from transit observations.

The temperatures for these 7 dwarfs were obtained in the same IRFM
implementation. However, with the exception of HD 209458, the
photometry was adopted from sources other than 2MASS, given that
they are bright stars. For $\tau$ Cet and GJ 105, TCS photometry
was available (Alonso et~al. 1994a). For Procyon and GJ380 we took
the mean of the Johnson photometry available in the GCPD. Finally,
for $\alpha$ Cen A and B, Johnson infrared magnitudes as observed
by Engels et al. (1981) were used. Johnson photometry was
transformed into the TCS system by using the transformation
equations of Alonso et~al. (1994a).

Very large systematic errors in the adopted $\fbol$ scale may lead
to a wrong $\teff$ scale. Our work adopts the fluxes as derived
from the AAM95 calibration, which is based on integrated UBVRIJHK
photometry and a theoretical correction for the flux outside the
U-K wavelength range. AAM95 showed their fluxes to be consistent
with previous works as that of Blackwell \& Petford (1991), which
was based on spectrophotometric data, at least for
solar-metallicity stars. They also showed that the observed
differences were obviously due to metallicity effects.

\begin{figure}
\plotone{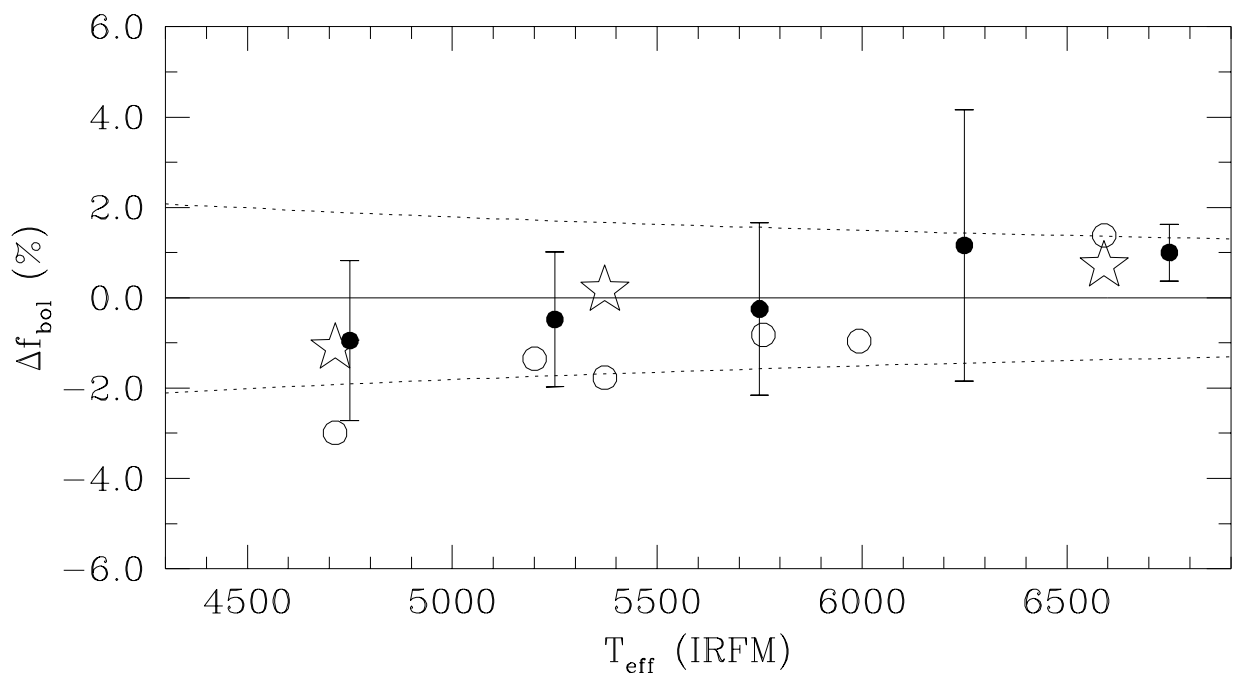}

\caption{Difference between the fluxes from the Alonso et~al.
(1995, AAM95) calibration and the fluxes from: Bessell et~al.
(1998) bolometric correction scale for the stars in
Table~\ref{table:diams} (open circles), Blackwell \& Lynas-Gray
(1998) measurements for 35 stars in 500~K bins (filled circles,
mean differences and $\pm$ 1 $\sigma$ error bars are shown), and
other measurements (see the text for references) for 3 stars in
Table~\ref{table:diams} (stars). The dotted lines illustrate how
an offset of $\pm30$~K in the derived IRFM $\teff$ scale may be
due to wrongly adopted bolometric fluxes.}

\label{fig:fbol}
\end{figure}

A comparison with the bolometric fluxes that are derived from the
theoretical BC scale of Bessell, Castelli, \& Plez (1998) for the
stars in Table \ref{table:diams} shows a tendency with
temperature. The fluxes from the theoretical BC's seem to be
higher by about 3\% at 4750~K but lower by about 1.5\%
at 6500~K (open circles in Fig.~\ref{fig:fbol}). 
Also, when comparing measured fluxes for 35 dwarf stars from the
Blackwell \& Lynas-Gray (1998) work with the measured fluxes used
by AAM95 to derive their calibration, the differences show a
similar behavior, in the sense that they increase with temperature
but only from $-$1\% to +1\%. The filled circles and error bars in
Fig.~\ref{fig:fbol} correspond to the mean difference and $\pm$ 1
$\sigma$ bars in 500~K bins, respectively. These small
discrepancies, however, do not affect our final results
substantially as will be explained in Sect.~\ref{sec:comparison}.

Measurements of $\fbol$ are only available for Procyon (Code
et~al. 1976, Beeckmans 1977, Blackwell \& Shallis 1977, Smalley \&
Dworetsky 1995, Mozurkewich et~al. 2003), $\tau$~Cet (Blackwell \&
Lynas-Gray 1994, 1998) and GJ~105~A (Blackwell \& Lynas-Gray
1998). By comparing the mean values of these measurements with the
$\fbol$ values we adopted, a mean difference of $(0.20\pm1.73)\%$
is found (the major part of the dispersion comes from GJ~105~A).
If we include the fluxes \textit{measured} by AAM95, the
difference reduces to $(0.08\pm0.94)\%$. Since there is no
apparent reason to exclude the AAM95 fluxes from the others, we
may conclude that the adopted $\fbol$ scale in our work is
consistent with the measurements and the dispersion is within the
adopted mean error for the fluxes. The stars in
Fig.~\ref{fig:fbol} show the difference between the flux as
derived from the AAM95 calibration and the mean of the
measurements.

The measured fluxes and the fluxes from the calibration of AAM95
are given in Table~\ref{table:diams}. They have been used to get
the \textit{direct} temperatures given in the same
Table~\ref{table:diams}, the measured fluxes have been preferred,
of course. For the 4 stars with no $\fbol$'s measured, we have
used the temperatures from the calibration. This is a reasonable
approximation given that the error in the flux is propagated to
$\sigma(\teff)$ roughly as $\sigma(\fbol)/4$ and that there is
evidence for the calibration to be in good agreement with the
measured fluxes.

The stars in Table \ref{table:diams} cover the range of
temperatures from 4000~K to 6600~K, which is approximately the
range covered by the planet-hosting stars. The zero point of the
IRFM $\teff$ scale for the dwarf stars is in good agreement with
the direct temperatures. Considering the best five measurements in
Table~\ref{table:diams}, the mean difference
($\teff^\mathrm{IRFM}-\teff^\mathrm{dir}$) is $-$14~K, with a
standard error of 24~K. Note that the angular diameters measured
for GJ~105~A and HD~209458 have large errors and so they are not
useful to constraint any $\teff$ scale. In fact, within
$\pm200$~K, any existing $\teff$ scale agree with these
measurements.

The temperatures derived in this work are given in Table
\ref{table:all}. Also given in Table \ref{table:all} are the radii
of these stars in solar units, obtained from their Hipparcos
parallaxes, the bolometric flux scale of AAM95 and the IRFM
angular diameters.

A $\teff$ scale for the planet-hosting stars is obtained by using
the $(V-K)$ color index, where $K$ is from 2MASS:
\begin{equation}
\frac{5040}{\teff}=0.460+0.313(V-K)-0.025(V-K)^2\ . \label{eq:cal}
\end{equation}
This formula was obtained from a fit to the data for the dwarf
stars with reliable 2MASS photometry in the list. Its standard
deviation is 43 K and its ranges of applicability are
$1.20<(V-K)<2.60$, $-0.50<\feh<0.45$. A plot of the data and
Eq.~(\ref{eq:cal}) is given in Fig.~\ref{fig:teffvk}a. Since the
metallicity range covered by these stars is short, the metallicity
dependence of this relation could not be distinguished from the
internal errors in the temperatures and so no additional terms
containing $\feh$ were included in the fit.

\section{Comparison with other studies} \label{sec:comparison}

\begin{figure}
\plotone{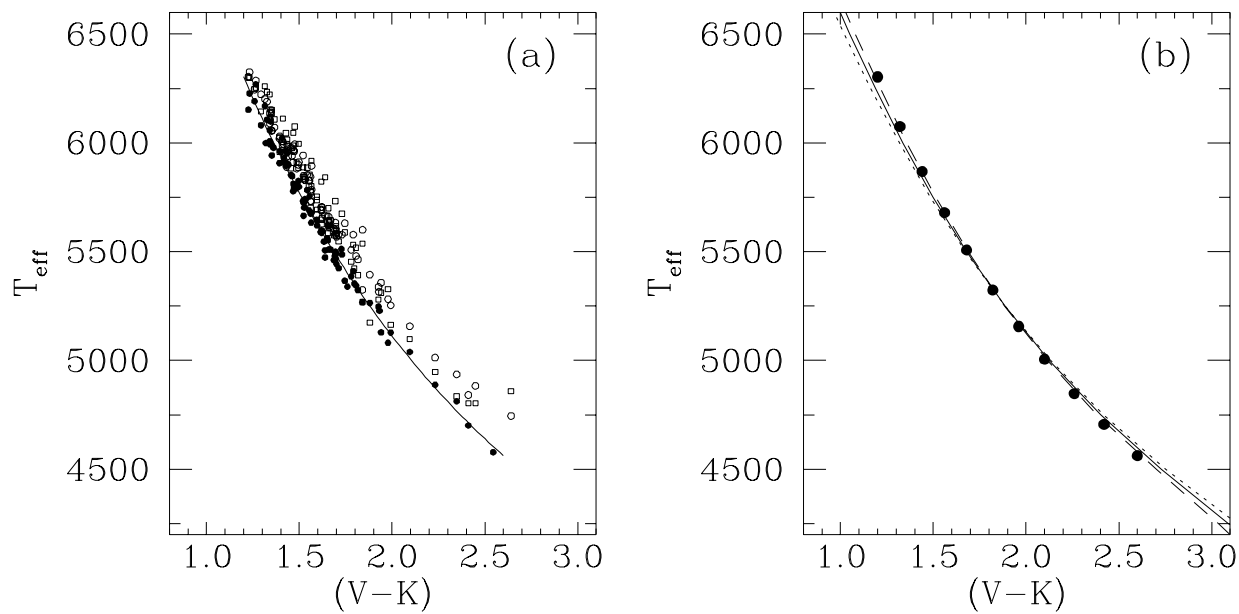} \caption{(a) $\teff$ vs $(V-K)$ for the
planet-hosting stars from the IRFM (filled circles), Ribas et~al.
(2003) (open circles) and Santos et al. (2004) (open squares).
Here, $K$ is from 2MASS and only stars with reliable 2MASS
photometry are plotted. The solid line corresponds to
Eq.~(\ref{eq:cal}). (b) $\teff$ vs $(V-K)$ for the colors derived
from Kurucz models for $\feh=+0.5$ (dotted line), $\feh=0$ (solid
line), and $\feh=-0.5$ (dashed line); the filled circles
correspond to Eq.~(\ref{eq:cal}).} \label{fig:teffvk}
\end{figure}

\begin{figure}
\plotone{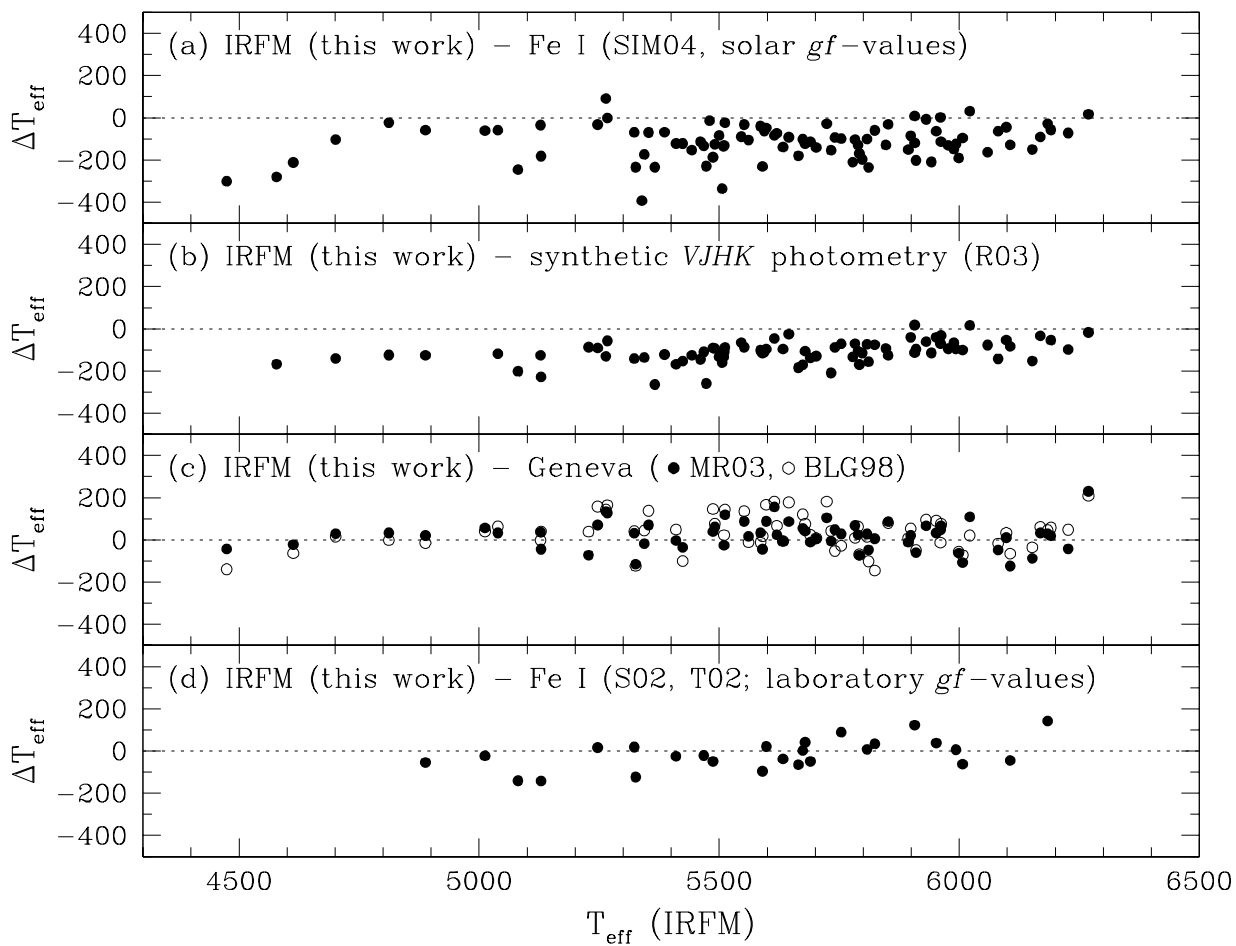}

\caption{Comparison between the temperatures obtained in this work
and those given by: (a) Santos et~al. (2004), (b) Ribas et~al.
(2003), (c) Mel\'endez \& Ram\'{\i}rez (2003) and Blackwell \&
Lynas-Gray (1998) IRFM calibrations for the Geneva colors, (d)
Sadakane et~al. (2002) and Takeda et~al. (2002).}

\label{fig:comparison}

\end{figure}

Our calibration for the $\teff$ vs $(V-K)$ relation is in
agreement with the colors derived from Kurucz models (M. Bessell,
private communication), as it is shown in Fig. \ref{fig:teffvk}b,
where the colors for $\feh=-0.5$ ,0.0 and $+0.5$ are plotted along
with our calibration. The colors kindly provided by M. Bessell are
in the Bessell \& Brett (1988) system and have been transformed
into 2MASS colors by using a transformation equation given by
Carpenter (2001). Although it is not shown in the figure, a good
agreement is also found with the solar metallicity $\teff$ vs
$(V-K)$ relation of Houdaschelt, Bell, \& Sweigart (2000), which
is based on MARCS models. On the other hand, the temperatures
given by R03 and SIM04 lead to a $\teff$ scale that has to be
shifted by about 110~K to better agree with them (squares and open
circles in Fig.~\ref{fig:teffvk}a). Note that a similar difference
($\teff^\mathrm{spec}-\teff^\mathrm{IRFM}=+139$~K) was found by
SIM04. Fig. \ref{fig:comparison}a and \ref{fig:comparison}b show
the differences between our $\teff$'s and those given by SIM04 and
R03, respectively. The calculated mean differences are
$-$115$\pm83$~K (91 stars) for the points in
Fig.~\ref{fig:comparison}a and $-108\pm53$~K (79 stars) for
Fig.~\ref{fig:comparison}b. The label `solar $gf$-values' in
Fig.~\ref{fig:comparison}a emphasizes the fact that SIM04 used
these transition probabilities in their analysis, while `synthetic
VJHK photometry' in Fig.~\ref{fig:comparison}b stands for the
basic idea of the R03 method.

The adopted $\fbol$ scale in our work can not be the only cause of
the difference, and, in fact, we believe it is not even the main
cause of it. 
The dotted lines in Fig.~\ref{fig:fbol} illustrate how an offset
in the derived IRFM temperatures corresponds to a systematic error
in the adopted bolometric fluxes. They have been derived by
assuming that the $\fbol$'s are the only source of error in the
implementation and so have produced temperatures systematically
lower and higher by 30~K compared to the absolute scale. The
quantities $\Delta\fbol$ are then the differences between the
adopted $\fbol$'s and the real ones. Even if the AAM95 calibration
is inaccurate, its effect on the $\teff$'s can not be larger than
about 30~K, and so it is very unlikely for it to be the main cause
of the 110~K difference.


The temperatures obtained from photometric calibrations based on
previous IRFM studies (e.g. Blackwell \& Lynas-Gray 1998;
Mel\'endez \& Ram\'{\i}rez 2003) agree very well with the present
results. As an illustration, the IRFM $\teff$ calibrations for the
Geneva colors $(B_2-G)$, $(B_2-V_1)$ and the $t$ parameter from
Mel\'endez \& Ram\'{\i}rez (2003, MR03) were applied and averaged
to get the $\teff$'s and compare them with the present results in
Fig.~\ref{fig:comparison}c. Also shown in
Fig.~\ref{fig:comparison}c is the difference between the $\teff$'s
obtained from the Blackwell \& Lynas-Gray (1998, BLG98) IRFM
$\teff$ calibration for $(B_2-V_1)$ and our
$\teff$'s. 

Not all the temperatures obtained from the excitation equilibrium
of Fe~I lines are in disagreement with the IRFM. The temperatures
obtained by Sadakane et~al. (2002, S02) and Takeda et~al. (2002,
T02), for instance, are in reasonable agreement with the IRFM
although a slight tendency with temperature is found
(Fig.~\ref{fig:comparison}d), the later maybe due to non-LTE
effects. This may also be the cause  of the small slope observed
in Fig. \ref{fig:comparison}a, which is roughly similar to that in
Fig. \ref{fig:comparison}d. The main difference with the work of
SIM04 seems to be the use of transition probabilities measured in
laboratory instead of solar $gf$-values. There is also a larger
coverage in excitation potential of Fe~I lines more sensitive to
temperature, SIM04 use 3 lines with $\chi_\mathrm{exc}<2$~eV
whilst S02 and T02 use 12. 
Only two of these lines are in the blue region of the spectrum,
where the continuum can not be accurately defined. However, all
the lines employed by S02 and T02, in addition to have reliable
experimental transition probabilities, were carefully selected and
so are hardly affected by blending.



\section{Conclusions}

The temperatures obtained in this work for the stars with planets
differ from those given by most of the other groups by about
110~K. In particular, there is a discrepancy with the excitation
equilibrium temperatures of Santos et al. (2004) 
and the method
of infrared photometry of Ribas et~al. (2003).

On the other hand, the $\teff$ vs $(V-K)$ relation from model
atmospheres is closer to our $\teff$ scale and the angular
diameters measured by interferometry and transit observations for 7
dwarf stars are well scaled to those derived from the IRFM.
The bolometric flux calibration adopted (AAM95) seems to be in
excellent agreement with the measurements reported in the
literature ($\pm1$\%), which strengthen the present results. The
comparison with direct measurements suggest that the zero point of
the IRFM $\teff$ scale is within 50~K, considering twice the
standard error. New measurements of angular diameters for stars in
the range 4000~K$<\teff<$6000~K are encouraged to better define
it.

The 110~K difference makes the stars bigger according to the IRFM,
which implies that the planetary radii obtained from transit
observations would be larger if our $\teff$ scale is adopted
instead of the spectroscopic one.  

Metallicity determinations for the stars with planets taking into
account the IRFM $\teff$ scale would be interesting. Note,
however, that large samples of stars both with and without
detected giant planets have been analyzed in a homogeneous way, so
it is unlikely that this will affect the well stated metallicity
enhancement of the planet-hosting stars, although small
differences may be found in an absolute scale.

\acknowledgments{We thank M. Bessell for providing colors for the
complete set of Kurucz models and I.~Ivans for her comments and
suggestions. Significant improvement to the original manuscript
was possible due to a very constructive refereeing process. 
This publication makes use of data
products from the Two Micron All Sky Survey, which is a joint
project of the University of Massachusetts and IPAC/Caltech,
founded by NASA and NSF.}

\clearpage

\begin{deluxetable}{crccccllc}

\tabletypesize{\scriptsize}

\tablecaption{Cool Dwarfs with Measured Angular Diameters}

\tablehead{ \colhead{Star name} & \colhead{HD}   &
\colhead{$\theta_\mathrm{LD}$ (mas)} &
\colhead{Ref.\tablenotemark{a}} &
\colhead{$\fbol$\tablenotemark{b}} &
\colhead{$\fbol$\tablenotemark{c}} & \colhead{$\teff^\mathrm{dir}$
(K)\tablenotemark{d}} & \colhead{$\teff^\mathrm{IRFM}$
(K)\tablenotemark{e}} & \colhead{$\theta_\mathrm{IRFM}$ (mas)}}

\startdata
Procyon & 61421 & 5.448 $\pm$ 0.052 & 1 & 1.837E$-5$ & 1.822E$-5$ & 6553 $\pm$ 40 & 6591 $\pm$ 73 & 5.405 $\pm$ 0.131 \\
$\tau$ Cet & 10700 & 1.971 $\pm$ 0.050 & 2 & 1.163E$-6$ & 1.154E$-6$ & 5473 $\pm$ 72 & 5372 $\pm$ 65 & 2.047 $\pm$ 0.054 \\
GJ 380 & 88230 & 1.306 $\pm$ 0.040\tablenotemark{f} & 3 & 1.373E$-7$ & \nodata & 3962 $\pm$ 63 & 3950 $\pm$ 161 & 1.301 $\pm$ 0.107 \\
GJ 105 A & 16160 & 0.941 $\pm$ 0.070 & 3 & 1.687E$-7$ & 1.725E$-7$ & 4917 $\pm$ 185 & 4714 $\pm$ 67 & 1.013 $\pm$ 0.031 \\
\nodata & 209458 & 0.226 $\pm$ 0.015 & 4 & 2.279E$-8$ & \nodata & 6049 $\pm$ 202 & 5993 $\pm$ 71 & 0.230 $\pm$ 0.006 \\
$\alpha$ Cen A & 128620 & 8.511 $\pm$  0.020 & 5 & 2.677E$-5$ & \nodata & 5771 $\pm$ 23 & 5759 $\pm$ 70 & 8.548 $\pm$ 0.225 \\
$\alpha$ Cen B & 128621 & 6.001 $\pm$  0.034 & 5 & 8.653E$-6$ &
\nodata & 5182 $\pm$ 24 & 5201 $\pm$ 65 & 5.957 $\pm$ 0.160
\enddata

\tablenotetext{a}{(1) Kervella et~al. (2004); (2) Pijpers et~al.
(2003); (3)Lane et~al. (2001); (4) Brown et~al. (2001); (5)
Kervella et~al. (2003).} \tablenotetext{b}{Bolometric fluxes from
the Alonso et~al. (1995) calibration. Units are erg cm$^{-2}$
s$^{-1}$.} \tablenotetext{c}{Mean of the measured bolometric
fluxes (see the text for references). Units are erg cm$^{-2}$
s$^{-1}$.} \tablenotetext{d}{Direct temperatures calculated by
using the measured bolometric fluxes or the adopted flux
calibration. An error of 1.5\% was adopted for the $\fbol$
values.} \tablenotetext{e}{Temperatures from the present IRFM
implementation.} \tablenotetext{f}{Estimated from the UD diameter
with a correction of 3\%.}

\label{table:diams}

\end{deluxetable}

\clearpage

\clearpage

\clearpage

\begin{deluxetable}{cccccclcc}

\tabletypesize{\scriptsize}

\tablecaption{Temperatures and Radii of the Planet-Hosting Stars
from the IRFM}

\tablehead{ \colhead{HD} & \colhead{$\teff$ (K)} & $R/R_\odot$ &
\colhead{HD} & \colhead{$\teff$ (K)} & $R/R_\odot$ &
\colhead{HD/BD} & \colhead{$\teff$ (K)} & $R/R_\odot$}

\startdata

000142   &   6152    $\pm$   77  &   1.467   $\pm$   0.032   &   065216   &   5561    $\pm$   66  &   0.900   $\pm$   0.036   &   143761   &   5754    $\pm$   70  &   1.341   $\pm$   0.029   \\
001237   &   5512    $\pm$   68  &   0.832   $\pm$   0.028   &   068988   &   5778    $\pm$   70  &   1.209   $\pm$   0.062   &   145675   &   5129    $\pm$   67  &   1.022   $\pm$   0.030   \\
002039   &   5847    $\pm$   69  &   1.251   $\pm$   0.105   &   070642   &   5620    $\pm$   67  &   1.023   $\pm$   0.031   &   147513   &   5852    $\pm$   62  &   0.950   $\pm$   0.026   \\
003651   &   5264    $\pm$   67  &   0.844   $\pm$   0.029   &   072659   &   5798    $\pm$   69  &   1.481   $\pm$   0.059   &   150706   &   5792    $\pm$   69  &   0.964   $\pm$   0.030   \\
004203   &   5546    $\pm$   68  &   1.418   $\pm$   0.102   &   073256   &   5344    $\pm$   65  &   0.971   $\pm$   0.039   &   162020   &   4578    $\pm$   68  &   0.800   $\pm$   0.056   \\
004208   &   5586    $\pm$   66  &   0.896   $\pm$   0.044   &   073526   &   5615    $\pm$   68  &   1.507   $\pm$   0.099   &   168443   &   5491    $\pm$   69  &   1.610   $\pm$   0.042   \\
006434   &   5741    $\pm$   70  &   1.077   $\pm$   0.045   &   074156   &   5910    $\pm$   70  &   1.667   $\pm$   0.076   &   168746   &   5468    $\pm$   69  &   1.148   $\pm$   0.049   \\
008574   &   5942    $\pm$   57  &   1.409   $\pm$   0.042   &   075289   &   6098    $\pm$   73  &   1.245   $\pm$   0.031   &   169830   &   6227    $\pm$   74  &   1.829   $\pm$   0.042   \\
009826   &   6184    $\pm$   75  &   1.613   $\pm$   0.028   &   075732   &   5247    $\pm$   69  &   0.934   $\pm$   0.030   &   177830   &   4701    $\pm$   67  &   3.354   $\pm$   0.054   \\
010697   &   5510    $\pm$   60  &   1.839   $\pm$   0.036   &   076700   &   5645    $\pm$   69  &   1.370   $\pm$   0.047   &   178911B  &   5366    $\pm$   69  &   1.195   $\pm$   0.233   \\
012661   &   5473    $\pm$   68  &   1.198   $\pm$   0.041   &   080606   &   5461    $\pm$   70  &   0.975   $\pm$   0.338   &   179949   &   6169    $\pm$   73  &   1.194   $\pm$   0.034   \\
013445   &   5128    $\pm$   56  &   0.797   $\pm$   0.025   &   082943   &   5952    $\pm$   71  &   1.147   $\pm$   0.035   &   186427   &   5633    $\pm$   71  &   1.190   $\pm$   0.029   \\
016141   &   5679    $\pm$   69  &   1.460   $\pm$   0.056   &   083443   &   5386    $\pm$   67  &   1.055   $\pm$   0.048   &   187123   &   5665    $\pm$   70  &   1.216   $\pm$   0.043   \\
017051   &   6269    $\pm$   66  &   1.085   $\pm$   0.025   &   089744   &   6106    $\pm$   71  &   2.182   $\pm$   0.037   &   190228   &   5081    $\pm$   65  &   2.648   $\pm$   0.057   \\
019994   &   5999    $\pm$   64  &   1.802   $\pm$   0.029   &   092788   &   5590    $\pm$   67  &   1.075   $\pm$   0.041   &   190360   &   5552    $\pm$   68  &   1.149   $\pm$   0.028   \\
020367   &   5989    $\pm$   72  &   1.197   $\pm$   0.039   &   106252   &   5907    $\pm$   70  &   1.098   $\pm$   0.043   &   192263   &   4888    $\pm$   65  &   0.781   $\pm$   0.036   \\
022049   &   5012    $\pm$   67  &   0.763   $\pm$   0.028   &   108147   &   6191    $\pm$   74  &   1.205   $\pm$   0.037   &   195019   &   5506    $\pm$   67  &   1.582   $\pm$   0.042   \\
023079   &   5961    $\pm$   71  &   1.121   $\pm$   0.032   &   108874   &   5443    $\pm$   69  &   1.262   $\pm$   0.089   &   196050   &   5789    $\pm$   69  &   1.287   $\pm$   0.050   \\
023596   &   5977    $\pm$   73  &   1.552   $\pm$   0.051   &   111232   &   5480    $\pm$   69  &   0.907   $\pm$   0.036   &   202206   &   5724    $\pm$   71  &   1.042   $\pm$   0.059   \\
027442   &   4613    $\pm$   67  &   4.179   $\pm$   0.032   &   114729   &   5783    $\pm$   68  &   1.480   $\pm$   0.043   &   209458   &   5993    $\pm$   71  &   1.165   $\pm$   0.054   \\
028185   &   5594    $\pm$   67  &   1.060   $\pm$   0.050   &   114762   &   5824    $\pm$   68  &   1.273   $\pm$   0.064   &   210277   &   5410    $\pm$   67  &   1.089   $\pm$   0.031   \\
030177   &   5500    $\pm$   69  &   1.157   $\pm$   0.050   &   114783   &   5039    $\pm$   66  &   0.804   $\pm$   0.035   &   213240   &   5899    $\pm$   70  &   1.545   $\pm$   0.042   \\
033636   &   5811    $\pm$   68  &   1.009   $\pm$   0.046   &   117176   &   5328    $\pm$   59  &   1.963   $\pm$   0.028   &   216435   &   5931    $\pm$   71  &   1.773   $\pm$   0.035   \\
037124   &   5424    $\pm$   67  &   1.033   $\pm$   0.047   &   121504   &   5962    $\pm$   70  &   1.140   $\pm$   0.048   &   216437   &   5733    $\pm$   63  &   1.514   $\pm$   0.029   \\
038529   &   5487    $\pm$   59  &   2.810   $\pm$   0.046   &   128311   &   4812    $\pm$   64  &   0.781   $\pm$   0.033   &   216770   &   5353    $\pm$   66  &   0.985   $\pm$   0.048   \\
039091   &   6022    $\pm$   65  &   1.121   $\pm$   0.025   &   130322   &   5323    $\pm$   64  &   0.808   $\pm$   0.052   &   217014   &   5690    $\pm$   61  &   1.162   $\pm$   0.026   \\
040979   &   6081    $\pm$   71  &   1.205   $\pm$   0.037   &   134987   &   5674    $\pm$   73  &   1.245   $\pm$   0.037   &   217107   &   5598    $\pm$   70  &   1.128   $\pm$   0.031   \\
046375   &   5267    $\pm$   67  &   1.021   $\pm$   0.045   &   136118   &   6059    $\pm$   70  &   1.758   $\pm$   0.051   &   219542B  &   5339    $\pm$   67  &   1.141   $\pm$   0.111   \\
049674   &   5509    $\pm$   70  &   0.987   $\pm$   0.054   &   137759   &   4474    $\pm$   64  &   12.494  $\pm$   0.034   &   222582   &   5702    $\pm$   70  &   1.133   $\pm$   0.054   \\
050554   &   5907    $\pm$   70  &   1.140   $\pm$   0.041   &   141937   &   5808    $\pm$   69  &   1.068   $\pm$   0.044   &   $-10$ 3166 &   5228    $\pm$   67  &  \nodata             \\
052265   &   6007    $\pm$   71  &   1.284   $\pm$   0.035   &   142415   &   5894    $\pm$   70  &   1.029   $\pm$   0.039   &   \nodata    &   \nodata            &  \nodata             \\

\enddata

\label{table:all}

\end{deluxetable}

\clearpage

\clearpage

\end{document}